# Bright excitons in monolayer transition metal dichalcogenides: from Dirac cones to Dirac saddle points

Hongyi Yu[1], Guibin Liu[2,1], Pu Gong[1], Xiaodong Xu[3,4], and Wang Yao[1*]

[1] Department of Physics and Center of Theoretical and Computational Physics, The University of Hong Kong, Hong Kong, China
[2] School of Physics, Beijing Institute of Technology, Beijing 100081, China
[3] Department of Physics, University of Washington, Seattle, Washington, USA
[4] Department of Material Science and Engineering, University of Washington, Seattle, Washington, USA

[*]Correspondence to: wangyao@hku.hk

**Abstract:** In monolayer transition metal dichalcogenides, tightly bound excitons have been discovered with a valley pseudospin that can be optically addressed through polarization selection rules. Here, we show that this valley pseudospin is strongly coupled to the exciton center-of-mass motion through electron-hole exchange. This coupling realizes a massless Dirac cone with chirality index $I$=2 for excitons inside the light cone, i.e. bright excitons. Under moderate strain, the $I$=2 Dirac cone splits into two degenerate $I$=1 Dirac cones, and saddle points with a linear Dirac spectrum emerge in the bright exciton dispersion. Interestingly, after binding an extra electron, the charged exciton becomes a massive Dirac particle associated with a large valley Hall effect protected from intervalley scattering. Our results point to unique opportunities to study Dirac physics, with exciton's optical addressability at specifiable momentum, energy and pseudospin. The strain-tunable valley-orbit coupling also implies new structures of exciton condensates, new functionalities of excitonic circuits, and possibilities for mechanical control of valley pseudospin.

**Introduction**

The strong coupling of an internal degree of freedom with motion can give rise to quasiparticles with radically new behaviors. A seminal example is the Dirac cones in graphene and topological insulators where low energy carriers are found with linear dispersion and spin-momentum locking[1-4]. These massless Dirac fermions are drawing great interest for their scientific and technological importance. Besides the Dirac cone, other band geometries exhibiting a linearly dispersing Dirac spectrum are being discovered including the Dirac circle and Dirac arc[5], the Dirac node[6], the anisotropic Dirac cone[7], and the three-dimensional Dirac point[8]. The interest on Dirac particles also drives the extensive efforts to realize Dirac spectra by engineering artificial honeycomb lattices of electrons, atoms and photons[9-11].

An exciton is a composite particle comprising of an electron and a hole in hydrogen-like bound state, which plays key roles in optoelectronic phenomena and applications. An exciton can be interconverted with a photon under the constraint of energy momentum conservation. In two-dimensional (2D) systems, this coupling to photons makes possible the optical preparation of exciton at any specified energy ($E$) and wavevector (**k**) on its dispersion in the *light cone* (the conical region defined by $E \geq c|\mathbf{k}|$), as well as the optical detection for mapping out the distribution of such *bright* excitons in energy-momentum space[10,11]. In semiconductors, the bright excitons can have lifetimes long enough to allow for the formation of an exciton condensate[12,13], and controlled flow in an integrated excitonic circuit[14].

Monolayer group-VIB transition metal dichalcogenides (TMDCs) offer a new 2D material system to explore exciton physics. These are direct bandgap semiconductors, with the conduction and valence band edges at the doubly degenerate corners (±K points) of the hexagonal Brillouin zone[15,16]. The bright exciton thus has two valley pseudospin configurations where the electron and hole are both confined at either the K or -K valley[17]. An optical selection rule dictates that an exciton in these two valley states couples respectively to a $\sigma+$ or $\sigma-$ circularly polarized photon only[18], which makes possible optical generation of excitonic valley polarization and coherence as demonstrated recently[19-22]. Moreover, the Coulomb interaction between the electron and hole is exceptionally strong due to their heavy masses and the 2D confinement. Sharp exciton and trion resonances are seen in the photoluminescence spectrum with a charging energy of ~ 30 meV[22-24], compatible with the exciton binding energy of ~ 0.5 − 1 eV from *ab initio* calculations[17,25-28].

Here we show that in monolayer TMDC the pronounced electron-hole Coulomb exchange in exciton realizes a strong valley-orbital coupling that can be orders of magnitude larger than the radiative recombination and momentum scattering rates. In such a *strong coupling regime*, the pseudospin splitting from the valley-orbit coupling becomes spectrally resolvable. A bright exciton in the light cone thus behaves like a massless Dirac particle, where the in-plane rotation symmetry dictates that it must have chirality index (*I*) of 2 (i.e. the valley pseudospin rotates $4\pi$ when momentum circles the Dirac point once), distinct from all known Dirac particles. Moderate tensile strain can realize a sizable in-plane Zeeman field on the exciton valley pseudospin, which splits the $I = 2$ Dirac cone into two degenerate $I = 1$ cones centered at opposite momentums. Consequently, a new type of band geometry, the linearly dispersing Dirac saddle points, emerges in the light cone. When the exciton binds an electron to form a trion, the exchange interaction with the excess electron opens up a gap at the Dirac point, turning the trion into a massive Dirac particle associated with a large valley Hall effect protected from intervalley scattering.

**Results**

**Massless Dirac cone of bright exciton**. The bright exciton $X_0$ in the two valley configurations is schematically shown in Fig. 1a. The valence band has a large valley-dependent spin splitting of ~ 150 − 450 meV[18,29]. The spin index of the band edge hole states are therefore locked with the valley index, i.e. K (-K) valley has spin up (down) states only. Spin is unchanged by the optical transition, leaving only two possible configurations for bright exciton. We use $\hat{\boldsymbol{\sigma}}$ to denote the exciton valley pseudospin, where pseudospin up (down) corresponds to $X_0$ at the K (-K) valley that couples to a $\sigma +$ ($\sigma -$) photon only. $X_0$ with an in-plane pseudospin, i.e. a superposition of the two valley configurations, then couples to a linearly polarized photon.

The Coulomb exchange interaction between the electron and hole of $X_0$ will couple its two valley configurations, giving rise to an effective interaction between the exciton's center-of-mass momentum $\mathbf{k} = (k\cos\theta, k\sin\theta)$ and its valley pseudospin $\hat{\boldsymbol{\sigma}}$: $\hat{H}_{\text{voc}} = J(\mathbf{k})\hat{\sigma}_+ + \text{c.c.}$ (see supplementary information). The valley-orbit coupling is of the form: $J(\mathbf{k}) = j(k)e^{-2i\theta}$, where the strength of the coupling $j(k) = -|\psi(0)|^2 \frac{a^2 t^2}{E_g^2} k^2 V(k)$. Here $a$ is the lattice constant, $E_g$ the bandgap, and $t$ is the nearest neighbor hopping integral in the monolayer TMDCs[18]. $V(k)$ is the Fourier transform of the Coulomb interaction, and $\psi(\mathbf{r}_{\text{eh}})$ is the wavefunction of the electron-hole relative motion. $|\psi(0)|^2$ thus corresponds to the probability of finding the electron and hole spatially overlapped. $|\psi(0)|^2 \sim a_B^{-2}$ where $a_B$ is the Bohr radius.

By this effective valley-orbit coupling, the exciton dispersion splits into two branches with energies $E_\pm(\mathbf{k}) = \omega_0 + \frac{k^2}{2M_0} \pm j(k)$ (see Fig. 1b). Each branch has a chirality index $I = 2$, i.e. the valley pseudospin lies in the plane and rotates twice when the momentum undergoes one full circle around $\mathbf{k} = 0$. This pseudospin chirality is dictated by the in-plane rotation symmetry of the lattice and the optical polarization selection rule associated with the exciton valley pseudospin. A photon emitted by the upper (lower) branch is linearly polarized with the projection of the polarization vector in the 2D plane being longitudinal (transverse) to the momentum $\mathbf{k}$. Such a longitudinal-transverse splitting is known in GaAs quantum wells[30], where it can give rise to exciton spin relaxation[31] and the optical spin Hall effect[32]. However, two features distinguish this effect in monolayer TMDCs. First, the exciton eigenstates are now coherent superpositions at two well-separated valleys in momentum space, where the splitting corresponds to a Rabi oscillation between the two valleys. Second, the magnitude of the splitting can be orders larger in TMDC monolayers than that in a GaAs quantum well, due to the much stronger Coulomb binding (represented by $|\psi(\mathbf{r}_{\text{eh}} = 0)|^2$)[23-25].

The Coulomb interaction in an intrinsic 2D semiconductor is of the unscreened form: $V(k) = \frac{2\pi e^2}{\epsilon k}$ [30]. Thus the valley-orbit coupling strength scales linearly with the momentum $k$: $j(k) = -\frac{k}{K} J$, where $J \approx 2\pi \frac{e^2}{\epsilon a_B} \left(\frac{t}{E_g}\right)^2 \frac{a}{a_B} (Ka)$. $K \equiv \frac{4\pi}{3a}$ is the distance from the $\pm K$ points to the $\Gamma$ point of the Brillouin zone. We note that $\frac{e^2}{\epsilon a_B}$ corresponds to the exciton binding energy which in monolayer TMDCs is $\sim 0.5 - 1$ eV[17,23-25], more than one order of magnitude larger than that in GaAs quantum wells. *Ab initio* calculations also find the exciton Bohr radius $a_B \sim 1$ nm[17,24,25,33], one order of magnitude smaller than that in GaAs. Thus, the splitting in monolayer

TMDCs is two orders of magnitude stronger than that in GaAs quantum wells. With the parameters $a$, $E_g$ and $t$ from band structure calculations[18], we estimate $J \sim 1$ eV.

Because of the strong valley-orbit coupling, the energy minimum of the exciton dispersion appears on a ring with radius $k_{\min} \sim 0.05K$, which is outside the light cone defined by $|\mathbf{k}| \leq \omega_0/c \sim 10^{-3}K$ (c.f. Fig. 1b). At the light cone edge, the valley pseudospin splitting $\frac{2\omega_0}{cK}J \sim 2$ meV, well exceeding the exciton's radiative decay rate and momentum scattering rate expected in clean samples at low temperature[13,19,34]. Thus, in the light cone, the exciton has a conical dispersion that can be spectrally resolved. The bright exciton behaves effectively like a massless Dirac particle, with a chirality index $I = 2$ and a group velocity $J/\hbar K \sim 10^5$ m/s. These are in sharp contrast to GaAs quantum wells, where the longitudinal-transverse splitting cannot be spectrally resolved[30,31].

The optical addressability of bright excitons provides unique opportunities to investigate Dirac particles. As an example, Fig. 1c schematically illustrates a setup for the direct observation of spin-momentum locking of massless Dirac particle. Using laser pulses with a narrow frequency and diffraction limited beam spot, we can resonantly excite excitons, selectively on an energy contour in the upper or lower branch of the Dirac cone. By using linearly polarized excitation[22], the population distribution of injected excitons is anisotropic in momentum space with the angular dependence of $\propto \cos^2\theta$ ($\propto \sin^2\theta$) if the upper (lower) branch is on resonance (c.f. Fig. 1c). Ballistic propagation will convert this angular distribution in momentum space into the same angular distribution in real space, which may be observed from the spatial map of the photoluminescence.

**Effects of strain on bright exciton and the Dirac saddle points**. The electron-hole exchange also realizes the coupling between exciton valley pseudospin and mechanical strain. Due to rotational symmetry of the monolayer TMDCs, the valley pseudospin splitting must vanish at $\mathbf{k} = 0$. However, an in-plane uniaxial tensile strain that breaks the rotational symmetry can introduce a finite splitting at $\mathbf{k} = 0$. Our *ab initio* calculation on monolayer WSe$_2$ shows that the strain realizes an in-plane Zeeman field on the valley pseudospin (see supplementary information), and the exciton Hamiltonian becomes

$$\widehat{H}_{0,\text{strain}} = \omega_0 + \frac{\hbar^2 k^2}{2M_0} - \frac{k}{K}Je^{-2i\theta}\hat{\sigma}_+ - \frac{k}{K}Je^{2i\theta}\hat{\sigma}_- + \boldsymbol{J_0} \cdot \hat{\boldsymbol{\sigma}}. \qquad (1)$$

For strain applied along the armchair (*x*) and zigzag (*y*) directions (see Fig. 2a), we find $\boldsymbol{J_0}$ along the positive *x*-axis and the negative *x*-axis respectively. Its magnitude as function of the strain strength is shown in Fig. 2c. A moderate strain of 1% can already induce a valley pseudospin splitting of ~ 12 meV, spectrally resolvable with existing sample qualities[22].

The strain effect makes possible a controllable way to coherently rotate the exciton valley pseudospin at **k** = 0. An even more interesting consequence is the Dirac cone with *I* = 2 now splits into two Dirac cones with *I* = 1 each, as shown in Fig. 2. From Eq. (1), the two new Dirac points are $\pm\frac{J_0}{J}K(\cos\frac{\theta_0}{2}, \sin\frac{\theta_0}{2})$, $\theta_0$ being the direction angle of $\boldsymbol{J_0}$. With a strain of 1%, both new Dirac points are well outside the light cone. The two new Dirac cones are anisotropic, with the velocity in the armchair direction being half of that in the zigzag direction. In the light cone, the two Dirac cones merge in the form of saddle points in both the upper and lower branches of the dispersion. Remarkably, the saddle points are linearly dispersed along all directions. Such Dirac type saddle points are unique to excitons in monolayer TMDCs and have not been discussed in any system. This is a generic consequence when a Dirac cone of chirality index *I* = 2 is subject to an in-plane Zeeman field on the pseudospin.

**Gapped Dirac cone and valley Hall effect of negatively charged trion**. Now we turn to negatively charged trion X- where an electron-hole pair binds an excess electron. Interestingly, the Coulomb exchange with the excess electron opens up a gap at the excitonic Dirac point, turning trion into a massive Dirac particle associated with a large Berry curvature. We note that conduction band edges in the ±K valleys also feature a valley-dependent spin splitting. The size of the splitting ranges from several to several tens of meV, and there is an overall sign difference between molybdenum dichalcogenides and tungsten dichalcogenides[29]. We focus on monolayer WSe$_2$ as example, where the spin splitting ~ 30 meV and the spin configurations of the conduction band edges in the ±K valleys are schematically shown in Fig. 1a. X- has four ground state configurations that can recombine to emit photon, as shown in Fig. 3. With the ~ 30 meV conduction band spin splitting, the excess electron is in the lowest energy band where optical transition to the valance band edge is spin forbidden. With the excess electron in the K (-K) valley, X- has the center-of-mass wave vector $\boldsymbol{q}_{X-} \equiv \boldsymbol{K} + \boldsymbol{k}$ ($\boldsymbol{q}_{X-} \equiv -\boldsymbol{K} + \boldsymbol{k}$). In principle, X-with any **k** can recombine as the recoil of the excess electron can facilitate the momentum conservation, but the recombination rate decays here as $\sim \exp(-k^2/(0.01K)^2)$,[24] which we define as the trion brightness. Trion emission is thus expected mainly in the region $k \lesssim 0.01K$.

The four X- configurations can be distinguished by $\hat{\sigma}_z$: the valley index of the electron-hole pair that can recombine, and $\hat{s}_z$: the spin index of the excess electron. The former is associated with the circular polarization of the emitted photon. Like $X_0$, the exchange interaction between the pair of electron and hole in the same valley with opposite spin will couple the two valley configurations with identical $\hat{s}_z$. Due to the rotational symmetry of the lattice, the coupling is still of the form $J(\mathbf{k}) = -\frac{k}{K} J e^{-2i\theta}$, where $\mathbf{k} = (k \cos\theta, k \sin\theta)$. The coupling strength $J$ is about the same to that in the $X_0$ case. In addition, there is also exchange interaction between the excess electron with the electron-hole pair (Fig. 3), which only affects the two X- configurations where $\hat{s}_z$ and $\hat{\sigma}_z$ have the same sign (first and third one in Fig. 3) and raises their energy by an amount estimated to be $\delta \sim 6$ meV (see supplementary information)[22]. The X- Hamiltonian is written

$$\hat{H}_- = \omega_- + \frac{\hbar^2 k^2}{2M_-} - \frac{k}{K} J e^{-2i\theta} \hat{\sigma}_+ - \frac{k}{K} J e^{2i\theta} \hat{\sigma}_- + \frac{\delta}{2}(\hat{\sigma}_z \hat{s}_z + 1). \qquad (2)$$

The exchange interaction of the electron-hole pair with the excess electron thus acts as a Zeeman field (with sign dependence on $\hat{s}_z$) on the valley pseudospin $\hat{\sigma}_z$, which opens up a finite gap at the two Dirac points $\mathbf{q}_{X-} \equiv \pm \mathbf{K}$. X- therefore behaves like a massive Dirac fermion in the bright regions in the neighborhood of the Dirac points.

The gap opening of the Dirac cone gives rise to a large Berry curvature with a peak value $\sim 10^4$ Å$^2$ in the neighborhood of $\mathbf{q}_{X-} \equiv \pm \mathbf{K}$, three orders of magnitude larger than that of the band edge electron in monolayer TMDCs[18]. In an in-plane electric field, X- as a charged particle will acquire a transverse velocity proportional to the Berry curvature, i.e. a Hall effect. Near the Dirac points where the curvature has large value, the trion eigenstates have $\langle \hat{\sigma}_z \rangle \approx \pm 1$, hence the photon emission is circularly polarized. We note that the sign of the Berry curvature is independent of $\hat{s}_z$, but is correlated with the polarization of the emitted photon. The two X- configurations that will emit $\sigma+$ ($\sigma-$) photon will move towards the right (left) edge of the system (c.f. Fig. 3). This trion valley Hall effect may be detected by the spatially and polarization resolved photoluminescence. Typically, scattering between the states with opposite Berry curvature will suppress the Hall effect. Remarkably, such scattering is inefficient here as it requires intervalley flips of two or three electrons/holes (c.f. Fig. 3), and thus the valley Hall effect of X- trion is protected.

**Discussion**

Electrostatic screening may modify the Dirac spectrum of exciton in monolayer TMDCs. In two dimensions, the Thomas-Fermi screened Coulomb potential is of the form $V(k) = \frac{2\pi e^2}{\epsilon(k+k_{\text{TF}})}$, where $k_{\text{TF}}$ is the screening wavevector. From the form of the valley-orbit coupling, it is clear that a finite screening will turn the dispersion of neutral exciton $X_0$ from linear to quadratic in a region of size $k_{\text{TF}}$ around the Dirac point, while the chirality index remains unchanged. Fig. 4a shows the $X_0$ dispersions in the light cone under different values of $k_{\text{TF}}$. For $k_{\text{TF}} \ll \omega_0/c$, $X_0$ behaves like a massless chiral particle, while for $k_{\text{TF}} \geq \omega_0/c$, it will gradually cross to a massive chiral particle like the low energy electrons in bilayer graphene. We estimate that $k_{\text{TF}} = \omega_0/c$ corresponds to a doping density of $10^{10}$ cm$^{-2}$ at temperature of ~ 30 K. Under the tensile strain, the spectrum will always split into two $I = 1$ massless Dirac cones regardless of the value of $k_{\text{TF}}$ (c.f. Fig. 4c). Nevertheless, $k_{\text{TF}}$ determines the dispersion of the saddle points where the two massless Dirac cones merge. The saddle point dispersion is linear for $k_{\text{TF}} \ll \omega_0/c$. For the negatively charged trion X-, we find the screening can induce a hole in the Berry curvature distribution at the Dirac point, while the peak value of the curvature remains the same order of magnitude for $k_{\text{TF}} \leq 10\omega_0/c$ (c.f. Fig. 4b).

To summarize, in monolayer TMDCs, the pronounced Coulomb exchange interaction gives rise to strong valley-orbit coupling for the neutral exciton $X_0$ and the negatively charged trion X-. In the bright regions of the momentum space where radiative recombination can occur, these excitons offer a new venue to explore the physics of massless and massive Dirac particles. Remarkably, the energy and momentum conservation of exciton-photon inter-conversion allows excitons to be selectively addressed in the light cone at any specified momentum and energy, through the control of laser frequency and propagation direction. The valley optical selection rule in monolayer TMDCs also allows the optical preparation and detection of excitonic valley pseudospin[19-22]. These features of the exciton system in monolayer TMDCs makes possible unique opportunities to observe behaviors of Dirac particles, as compared to the electronic counterparts. Moreover, the large strain-tunable valley-orbit coupling makes monolayer TMDCs an extremely interesting playground to explore novel optoelectronic applications such as the excitonic circuit[14] and exotic quantum phenomena such as exciton condensation[12,13]. The ring geometry of the energy minimum and the pseudospin texture will likely lead to new structures of exciton condensates in monolayer TMDCs.

**Acknowledgments:** We thank Z. D. Wang for helpful discussions and J. Schaibley for proof reading. This work is mainly supported by the Croucher Foundation under the Croucher Innovation Award, and the Research Grant Council of Hong Kong (HKU705513P, HKU8/CRF/11G). X.X. is supported by US DoE, BES, Materials Sciences and Engineering Division (DE-SC0008145) and NSF (DMR-1150719).

**Competing Final Interests**

The authors declare no competing financial interests.

**Figures**

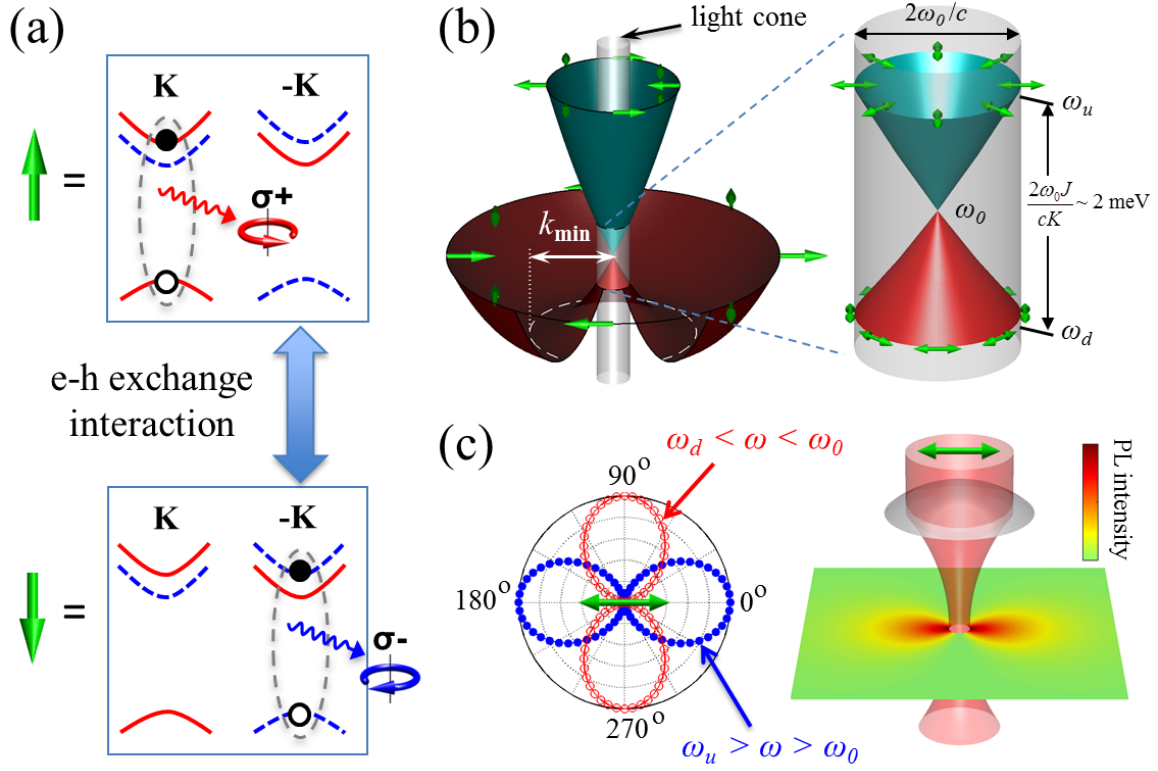

**Figure 1. Valley-orbit coupled neutral exciton $X_0$.** (a) Valley pseudospin-up and -down configurations of $X_0$, which recombine to emit $\sigma+$ and $\sigma-$ photon respectively. Red (blue) lines denotes spin up (down) conduction and valance band edges in K and –K valleys, and solid (hollow) circle denotes the electron (hole). Electron-hole exchange interaction acts as an in-plane field that couples the two valley configurations depending on the exciton center-of-mass wavevector **k**. (b) Dispersion of valley-orbit coupled $X_0$, which, in the light cone, realizes a massless Dirac cone with chirality index $I = 2$. The single-head arrows denote the valley pseudospin at various **k**. In the zoom-in of the dispersion in the light cone, double-head arrows are used to denote the linear polarization of the emitted photon upon exciton recombination. (c) Schematic setup for observing spin-momentum locking of $X_0$. Blue (red) dots in the polar plot denote the angular distribution of $X_0$ in k-space when the upper (lower) branch of Dirac cone is resonantly excited by a linearly polarized laser with a diffraction-limited beam focus. Ballistic transport converts this distribution in k-space into the same angular distribution in real space for observation.

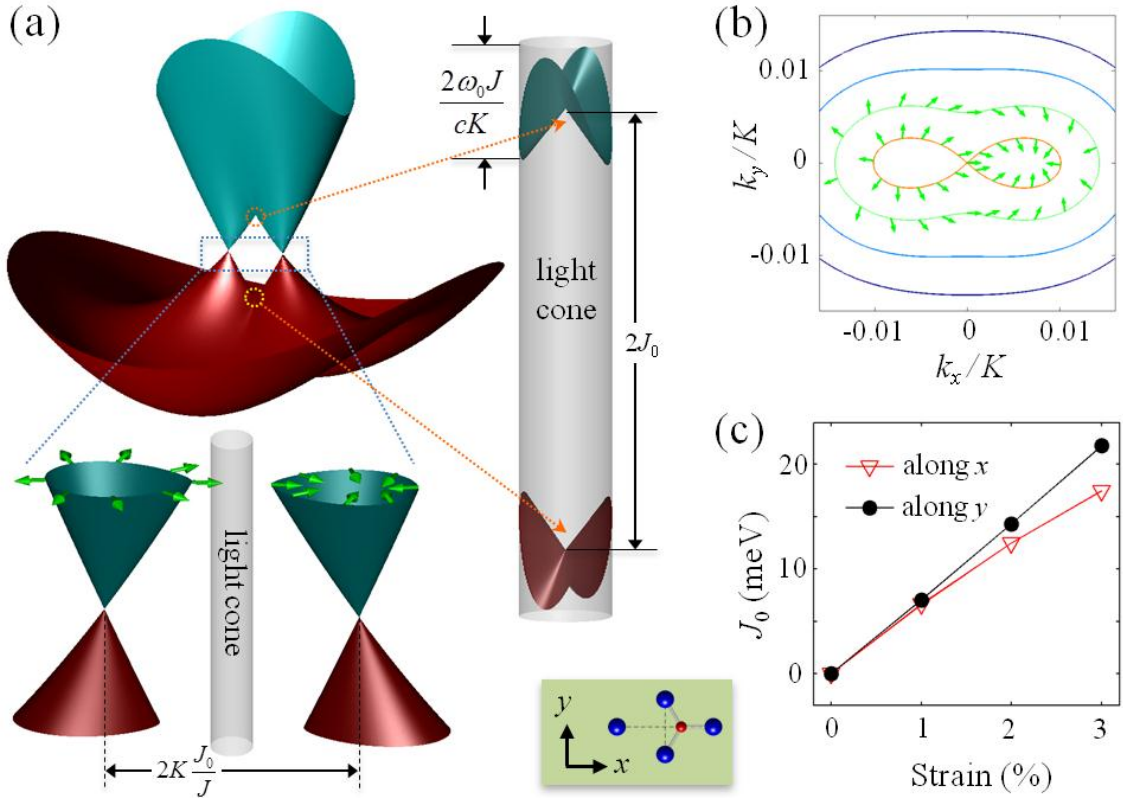

**Figure 2. Strain effect on $X_0$.** (a) The strain acts as an in-plane Zeeman field $J_0$ on valley pseudospin, splitting the single Dirac cone of chirality index $I = 2$ into two Dirac cones of $I = 1$. Consequently, two Dirac saddle points of linear dispersion appear in the light cone. (b) Contour plot of the higher energy branch. The green arrows illustrate the valley pseudospin configurations. (c) The calculated magnitude of $J_0$ as a function of the strength of strain applied along the $x$ and $y$ directions shown in the inset (see supplementary information).

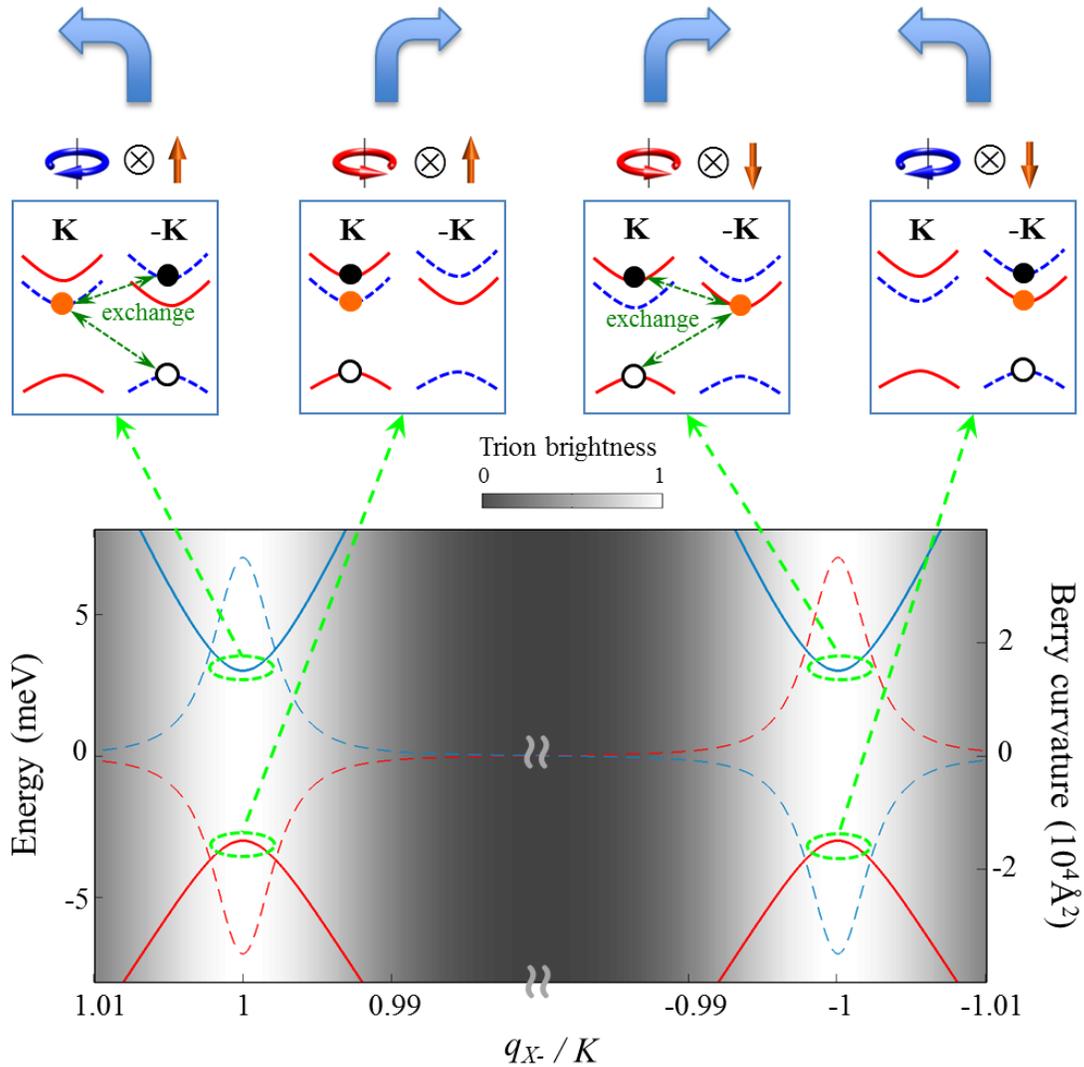

**Figure 3. Massive Dirac cones of negatively charged trion X-**. Upper: the four configurations of X- that can emit photons, labeled by the spin of the excess electron and polarization of the photon emission. The latter is determined by the valley pseudospin of the electron-hole pair that can recombine (black circles). Lower: the solid lines plot the dispersion of X-, where a gap is opened at the Dirac points by the exchange coupling of the electron-hole pair with the excess electron. Whiteness of the background corresponds to the brightness of X- as function of center-of-mass wavevector $\mathbf{q}_{X-}$. The blue (red) dashed lines plot the Berry curvature of the upper (lower) branch of the dispersion, which gives rise to the anomalous transverse motion of X- in an electric field, as illustrated by curved arrows on top.

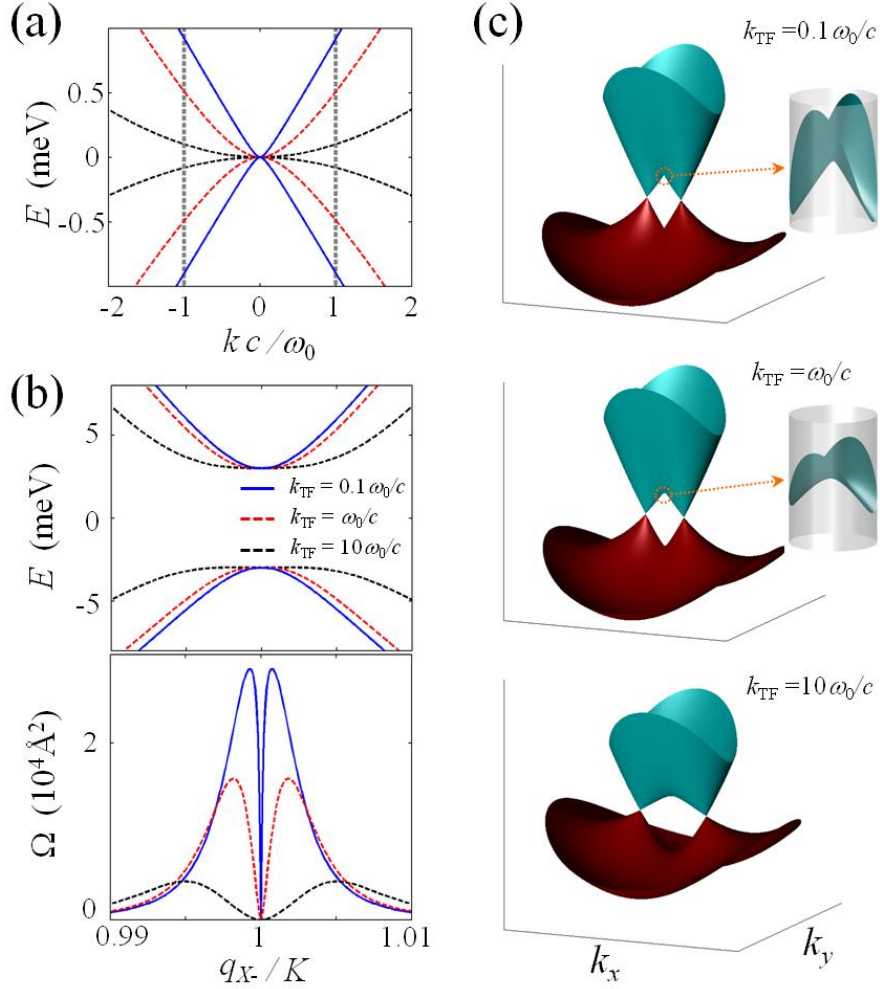

**Figure 4. Effect of Coulomb screening**. (a) Dispersion of $X_0$ in the light cone. (b) Dispersion and Berry curvature of X-. For the blue, red and black curves, the Thomas-Fermi screening wavevector $k_{TF} = 0.1$, 1 and 10 respectively, in units of $\omega_0/c$. (c) Dispersion of $X_0$ under 1% strain. For the values of $k_{TF}$ considered, the dispersions near the pair of new Dirac points are always linear. The saddle point in the light cone changes gradually from linear dispersion for $k_{TF} \ll \omega_0/c$ to quadratic one for $k_{TF} \geq \omega_0/c$.

# Bright excitons in monolayer transition metal dichalcogenides: from Dirac cones to Dirac saddle points (Supplementary Material)

Hongyi Yu, Guibin Liu, Pu Gong, Xiaodong Xu, and Wang Yao

## S1. Electron-hole exchange in the neutral exciton $X_0$

In monolayer transition metal dichalcogenides, at the valance band edges, the spin index is locked with the valley index, i.e. K (-K) valley has only spin down (up) holes. The electron-hole (e-h) exchange interaction can be written as [S1],

$$\hat{H}_{\text{eh-ex}} \equiv \sum_{\mathbf{q},\mathbf{k},\mathbf{k'}} J_{\text{ex}}(\mathbf{q},\mathbf{q'},\mathbf{k})\hat{h}^\dagger_{-\mathbf{K}-\mathbf{q'}+\frac{\mathbf{k}}{2},\Downarrow}\hat{e}^\dagger_{\mathbf{K}+\mathbf{q'}+\frac{\mathbf{k}}{2},\uparrow}\hat{e}_{-\mathbf{K}+\mathbf{q}+\frac{\mathbf{k}}{2},\downarrow}\hat{h}_{\mathbf{K}-\mathbf{q}+\frac{\mathbf{k}}{2},\Uparrow} + \text{h.c.}.$$

This term annihilates an electron-hole pair with center-of-mass wave vector $\mathbf{k}$ and relative wave vector $\mathbf{q}$ in the $-\mathbf{K}$ valley and creates a pair in the $\mathbf{K}$ valley with center-of-mass wave vector $\mathbf{k}$ and relative wave vector $\mathbf{q'}$, and vise versa. This process can also be viewed in the picture of conduction and valence states of the electrons as shown in Fig. S1. We note that such e-h exchange coupling only exists for bright excitons where the electron and hole have opposite spin.

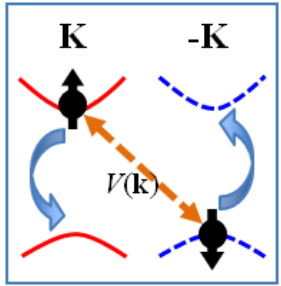

Fig. S1 The e-h exchange interaction depicted in terms of the conduction and valance states of the electrons. The conduction band electron in K valley is scattered to the valance state in the same valley, while the valance electron in –K valley is scattered to the conduction state in valley –K, as indicate by the curved blue arrows.

The matrix element

$$J_{\text{ex}}(\mathbf{q},\mathbf{q'},\mathbf{k}) = \int d\mathbf{r}_1 d\mathbf{r}_2 \psi^*_{\mathbf{K}+\mathbf{q'}+\frac{\mathbf{k}}{2},c,\uparrow}(\mathbf{r}_1)\psi_{\mathbf{K}+\mathbf{q'}-\frac{\mathbf{k}}{2},v,\uparrow}(\mathbf{r}_1)V(\mathbf{r}_1-\mathbf{r}_2)\psi_{-\mathbf{K}+\mathbf{q}+\frac{\mathbf{k}}{2},c,\downarrow}(\mathbf{r}_2)\psi^*_{-\mathbf{K}+\mathbf{q}-\frac{\mathbf{k}}{2},v,\downarrow}(\mathbf{r}_2).$$

Here $V(\mathbf{r})$ is the Coulomb potential. The Bloch wave functions:
$\psi_{\mathbf{K}+\mathbf{q},c(v),\uparrow}(\mathbf{r}) = e^{i(\mathbf{K}+\mathbf{q})\cdot\mathbf{r}} u_{\mathbf{K}+\mathbf{q},c(v),\uparrow}(\mathbf{r})$, and $\psi_{-\mathbf{K}+\mathbf{q},c(v),\downarrow}(\mathbf{r}) = e^{i(-\mathbf{K}+\mathbf{q})\cdot\mathbf{r}} u_{-\mathbf{K}+\mathbf{q},c(v),\downarrow}(\mathbf{r})$. Then the matrix element becomes

$$J_{\text{ex}}(\mathbf{q},\mathbf{q'},\mathbf{k}) = \sum_{\mathbf{G}} \frac{V(\mathbf{G}+\mathbf{k})}{A} \left\langle u_{-\mathbf{K}+\mathbf{q}-\frac{\mathbf{k}}{2},v,\downarrow} \left| e^{-i\mathbf{G}\cdot\mathbf{r}} \right| u_{-\mathbf{K}+\mathbf{q}+\frac{\mathbf{k}}{2},c,\downarrow} \right\rangle \left\langle u_{\mathbf{K}+\mathbf{q'}+\frac{\mathbf{k}}{2},c,\uparrow} \left| e^{i\mathbf{G}\cdot\mathbf{r}} \right| u_{\mathbf{K}+\mathbf{q'}-\frac{\mathbf{k}}{2},v,\uparrow} \right\rangle,$$

where $\mathbf{G}$ are reciprocal lattice vectors. $V(\mathbf{k}) \equiv \int V(\mathbf{r})e^{i\mathbf{k}\cdot\mathbf{r}}d\mathbf{r}$ denotes the Coulomb interaction in $\mathbf{k}$-space, and $A$ is the area of the 2D plane.

The creation operator for a bright exciton with center-of-mass wave vector $\mathbf{k}$ is $\hat{B}_{\mathbf{k},+}^{\dagger} \equiv \sum_{\mathbf{q}} \psi(\mathbf{q}) \hat{e}_{\mathbf{K}+\mathbf{q}+\frac{\mathbf{k}}{2},\uparrow}^{\dagger} \hat{h}_{-\mathbf{K}-\mathbf{q}+\frac{\mathbf{k}}{2},\Downarrow}^{\dagger}$ ($\hat{B}_{\mathbf{k},-}^{\dagger} \equiv \sum_{\mathbf{q}} \psi^*(-\mathbf{q}) \hat{e}_{-\mathbf{K}+\mathbf{q}+\frac{\mathbf{k}}{2},\downarrow}^{\dagger} \hat{h}_{\mathbf{K}-\mathbf{q}+\frac{\mathbf{k}}{2},\Uparrow}^{\dagger}$), where the subscript '+' ('-') denotes the exciton in valley $\mathbf{K}$ (-$\mathbf{K}$). $\psi(\mathbf{q})$ is the wavefunction for the relative motion between the electron and hole. The e-h exchange interaction will lead to the coupling between the two valley configurations of bright excitons: $\hat{H}_{\text{ex}} = J_{\mathbf{k}} \hat{B}_{\mathbf{k}+}^{\dagger} \hat{B}_{\mathbf{k}-} + \text{h.c.}$, where

$$J_{\mathbf{k}} = \sum_{\mathbf{qq}'} \psi(-\mathbf{q}) \psi(\mathbf{q}') J_{\text{ex}}(\mathbf{q}, \mathbf{q}', \mathbf{k})$$

$$= \sum_{\mathbf{G}} \frac{V(\mathbf{G}+\mathbf{k})}{A} \left( \sum_{\mathbf{q}} \psi(-\mathbf{q}) \left\langle u_{-\mathbf{K}+\mathbf{q}-\frac{\mathbf{k}}{2},v,\downarrow} \middle| e^{-i\mathbf{G}\cdot\mathbf{r}} \middle| u_{-\mathbf{K}+\mathbf{q}+\frac{\mathbf{k}}{2},c,\downarrow} \right\rangle \right)$$

$$\times \left( \sum_{\mathbf{q}'} \psi(\mathbf{q}') \left\langle u_{\mathbf{K}+\mathbf{q}'+\frac{\mathbf{k}}{2},c,\uparrow} \middle| e^{i\mathbf{G}\cdot\mathbf{r}} \middle| u_{\mathbf{K}+\mathbf{q}'-\frac{\mathbf{k}}{2},v,\uparrow} \right\rangle \right).$$

The summation over $\mathbf{G}$ can be separated into two parts: the long range part with $\mathbf{G} = 0$, and the short range part with $\mathbf{G} \neq 0$. The long range part is

$$J_{\mathbf{k}}^{\text{LR}} = \frac{V(\mathbf{k})}{A} \left( \sum_{\mathbf{q}} \psi(-\mathbf{q}) \left\langle u_{-\mathbf{K}+\mathbf{q}-\frac{\mathbf{k}}{2},v,\downarrow} \middle| u_{-\mathbf{K}+\mathbf{q}+\frac{\mathbf{k}}{2},c,\downarrow} \right\rangle \right)$$

$$\times \left( \sum_{\mathbf{q}'} \psi(\mathbf{q}') \left\langle u_{\mathbf{K}+\mathbf{q}'+\frac{\mathbf{k}}{2},c,\uparrow} \middle| u_{\mathbf{K}+\mathbf{q}'-\frac{\mathbf{k}}{2},v,\uparrow} \right\rangle \right).$$

To evaluate the matrix element $\left\langle u_{\mathbf{K}+\mathbf{q}+\frac{\mathbf{k}}{2},c,\uparrow} \middle| u_{\mathbf{K}+\mathbf{q}-\frac{\mathbf{k}}{2},v,\uparrow} \right\rangle$, we use the $\mathbf{k} \cdot \mathbf{p}$ expansion:

$$|u_{\mathbf{K}+\mathbf{q}+\frac{\mathbf{k}}{2},c}\rangle = |u_{\mathbf{K}+\mathbf{q},c}\rangle + \frac{\hbar}{2m} \mathbf{k} \cdot \sum_{n \neq c} \frac{\langle u_{\mathbf{K}+\mathbf{q},n} | \hat{\mathbf{p}} | u_{\mathbf{K}+\mathbf{q},c} \rangle}{E_{\mathbf{K}+\mathbf{q},c} - E_{\mathbf{K}+\mathbf{q},n}} |u_{\mathbf{K}+\mathbf{q},n}\rangle + O(k^2),$$

$$|u_{\mathbf{K}+\mathbf{q}-\frac{\mathbf{k}}{2},v}\rangle = |u_{\mathbf{K}+\mathbf{q},v}\rangle + \frac{\hbar}{2m} \mathbf{k} \cdot \sum_{n \neq v} \frac{\langle u_{\mathbf{K}+\mathbf{q},n} | \hat{\mathbf{p}} | u_{\mathbf{K}+\mathbf{q},v} \rangle}{E_{\mathbf{K}+\mathbf{q},n} - E_{\mathbf{K}+\mathbf{q},v}} |u_{\mathbf{K}+\mathbf{q},n}\rangle + O(k^2),$$

Then $\left\langle u_{\mathbf{K}+\mathbf{q}+\frac{\mathbf{k}}{2},c} \middle| u_{\mathbf{K}+\mathbf{q}-\frac{\mathbf{k}}{2},v} \right\rangle = \frac{\hbar}{m} \mathbf{k} \cdot \frac{\langle u_{\mathbf{K}+\mathbf{q},c} | \hat{\mathbf{p}} | u_{\mathbf{K}+\mathbf{q},v} \rangle}{E_{\mathbf{K}+\mathbf{q},c} - E_{\mathbf{K}+\mathbf{q},v}} + O(k^2)$. We are interested in the bright exciton within and near the light cone, i.e. the exciton center-of-mass wavevector $|\mathbf{k}| \sim \mu\text{m}^{-1}$ which is a small quantity compared to the size of the Brillouin zone. Thus, in the above equations, we only need to keep the leading term in $\mathbf{k}$. We note that $\mathbf{d}_{cv,\pm\mathbf{K}+\mathbf{q}} \equiv \frac{i\hbar}{m} \frac{\langle u_{\pm\mathbf{K}+\mathbf{q},c} | \hat{\mathbf{p}} | u_{\pm\mathbf{K}+\mathbf{q},v} \rangle}{E_{\pm\mathbf{K}+\mathbf{q},c} - E_{\pm\mathbf{K}+\mathbf{q},v}}$ is the optical dipole matrix element [S6]. The long range part of the exchange is then $J_{\mathbf{k}}^{\text{LR}} = V(\mathbf{k}) (\mathbf{k} \cdot \mathbf{d}_{X_0,+})(\mathbf{k} \cdot \mathbf{d}_{X_0,-}^*)$, where $\mathbf{d}_{X_0,+} \equiv \frac{1}{\sqrt{A}} \sum_{\mathbf{q}} \psi(\mathbf{q}) \mathbf{d}_{cv,\mathbf{K}+\mathbf{q}}$ ($\mathbf{d}_{X_0,-} \equiv \frac{1}{\sqrt{A}} \sum_{\mathbf{q}} \psi^*(-\mathbf{q}) \mathbf{d}_{cv,-\mathbf{K}+\mathbf{q}}$) is the optical matrix element of the exciton at valley K (-K) with zero center-of-mass wavevector. The valley optical selection rule [S2] require that exciton in valley K (-K) couples only to the $\sigma +$ ($\sigma -$) polarized photon, so $\mathbf{d}_{X_0,+} \equiv (d_{X_0,+}^x, d_{X_0,+}^y) = (d_{X_0,+}^x, -id_{X_0,+}^x) = \mathbf{d}_{X_0,-}^*$. Thus $J_{\mathbf{k}}^{\text{LR}} = V(\mathbf{k}) k^2 (d_{X_0,+}^x)^2 e^{-2i\theta}$, with $\mathbf{k} \equiv (k_x, k_y) = (k\cos\theta, k\sin\theta)$.

For an approximation, we can write $\mathbf{d}_{X_0,+} \equiv \frac{1}{\sqrt{A}} \sum_{\mathbf{q}} \psi(\mathbf{q}) \mathbf{d}_{cv,\mathbf{K}+\mathbf{q}} \approx \frac{\mathbf{d}_{cv,\mathbf{K}}}{\sqrt{A}} \sum_{\mathbf{q}} \psi(\mathbf{q})$. From the two-band $\mathbf{k}.\mathbf{p}$ Hamiltonian in Ref. [S2], we obtain $\mathbf{d}_{cv,\mathbf{K}} = \left( i\frac{at}{E_g}, \frac{at}{E_g} \right)$, where $a$ the lattice constant of monolayer TMDCs, $t$ the hopping amplitude, and $E_g$ the band gap which can all be fitted from the *ab initio* band structures. Then

$$J_{\mathbf{k}}^{\mathrm{LR}} = -\frac{|\sum_{\mathbf{q}} \psi(\mathbf{q})|^2}{A} \frac{a^2 t^2}{E_g^2} V(\mathbf{k}) k^2 e^{-2i\theta} = -|\psi(\mathbf{r}_{\mathrm{eh}} = 0)|^2 \frac{a^2 t^2}{E_g^2} V(\mathbf{k}) k^2 e^{-2i\theta}.$$

Here $\psi(\mathbf{r}_{\mathrm{eh}}) \equiv \frac{1}{\sqrt{A}} \sum_{\mathbf{q}} \psi(\mathbf{q}) e^{-i\mathbf{q} \cdot \mathbf{r}_{\mathrm{eh}}}$ is the real space wavefunction for the relative motion between electron and hole. $|\psi(\mathbf{r}_{\mathrm{eh}} = 0)|^2$ thus corresponds to the probability density of finding the electron and hole to spatially overlap, which is $\sim 1/a_B^2$ with $a_B$ the exciton Bohr radius.

The short range part of the exchange is

$$J_{\mathbf{k}}^{\mathrm{SR}} = \sum_{\mathbf{G} \neq 0} \frac{V(\mathbf{G} + \mathbf{k})}{A} \left( \sum_{\mathbf{q}} \psi(-\mathbf{q}) \left\langle u_{-\mathbf{K}+\mathbf{q}-\frac{\mathbf{k}}{2},v,\downarrow} \left| e^{-i\mathbf{G} \cdot \mathbf{r}} \right| u_{-\mathbf{K}+\mathbf{q}+\frac{\mathbf{k}}{2},c,\downarrow} \right\rangle \right)$$
$$\times \left( \sum_{\mathbf{q}'} \psi(\mathbf{q}') \left\langle u_{\mathbf{K}+\mathbf{q}'+\frac{\mathbf{k}}{2},c,\uparrow} \left| e^{i\mathbf{G} \cdot \mathbf{r}} \right| u_{\mathbf{K}+\mathbf{q}'-\frac{\mathbf{k}}{2},v,\uparrow} \right\rangle \right).$$

We write $J_{\mathbf{k}}^{\mathrm{SR}} = J_{\mathbf{k}=0}^{\mathrm{SR}} + O(k)$. Since $|\mathbf{k}| \ll |\mathbf{G}|$, we have $V(\mathbf{G} + \mathbf{k}) \ll V(\mathbf{k})$, the magnitude of the $O(k)$ term in $J_{\mathbf{k}}^{\mathrm{SR}}$ is much smaller than that in $J_{\mathbf{k}}^{\mathrm{LR}}$. Below, we examine the $J_{\mathbf{k}=0}^{\mathrm{SR}}$ term using symmetry considerations. With a nonzero value of $J_{\mathbf{k}=0}^{\mathrm{SR}}$, the split exciton eigenstates at $\mathbf{k} = 0$ will be linear combinations of the two valleys, which will emit linear polarized photons upon recombination. A nonzero $J_{\mathbf{k}=0}^{\mathrm{SR}}$ thus corresponds to a longitudinal-transverse splitting of the photon emission at $\mathbf{k} = 0$ which violates the three fold rotational symmetry of the lattice. This rotation symmetry dictates $J_{\mathbf{k}=0}^{\mathrm{SR}} = 0$. Therefore, in the presence of rotation symmetry, the e-h exchange induced valley coupling is: $J_{\mathbf{k}} \approx J_{\mathbf{k}}^{\mathrm{LR}} = -\frac{a^2 t^2}{a_B^2 E_g^2} V(\mathbf{k}) k^2 e^{-2i\theta}$. The dependence of the valley coupling on the exciton center-of-mass wavevector $\mathbf{k}$ means that the e-h exchange manifests itself as an effective valley-orbital coupling.

If the 2D Coulomb interaction is of the unscreened form $V(\mathbf{k}) = \frac{2\pi e^2}{\epsilon k}$, then $J_{\mathbf{k}}$ scales linearly with $k$. In a doped semiconductor, we consider the Thomas Fermi screening of the 2D Coulomb potential: $V(\mathbf{k}) = \frac{2\pi e^2}{\epsilon(k+k_{\mathrm{TF}})}$. For small $k$, the screening wavevector has the form $k_{\mathrm{TF}} = \frac{2m^* e^2}{\epsilon \hbar^2} f_0$ [S6], where $f_0$ is the Fermi–Dirac distribution of the electron and hole at the $\pm \mathbf{K}$ point. As an example, consider a temperature of 30 K, with the carrier density of n $\sim 10^9$ cm$^{-2}$, we find $k_{\mathrm{TF}} \approx 0.1 \, \omega_0/c$. For n $\sim 10^{10}$ cm$^{-2}$, $k_{\mathrm{TF}} \approx \omega_0/c$, and for n $\sim 10^{11}$ cm$^{-2}$ we have $k_{\mathrm{TF}} \approx 10 \omega_0/c$, where $\omega_0/c$ is the size of the light cone.

With the unscreened Coulomb potential, the valley-orbit coupling is therefore $J_{\mathbf{k}} \approx -\frac{a^2 t^2}{a_B^2 E_g^2} \frac{2\pi e^2}{\epsilon} k e^{-2i\theta} = -J \frac{k}{K} e^{-2i\theta}$, where $J = 2\pi \frac{e^2}{\epsilon a_B} \left(\frac{t}{E_g}\right)^2 \frac{a}{a_B} (Ka)$. Here $K \equiv 4\pi/3a$ is the distance from the K to $\Gamma$ point, which gives the size of the Brillouin zone. The relevant parameters for the four group VIB transition metal dichalcogenides monolayers are in the same order of magnitude [S2]. Below we use the ones for monolayer WSe$_2$. The parameters are: $a = 3.31$ Å, $K = 4\pi/3a = 1.26$ Å$^{-1}$, $t = 1.19$ eV, $E_g = 2$ eV. Then we have,

$$J = 1 \text{ eV} \times \frac{E_b/500 \text{ meV}}{a_B/1 \text{ nm}}.$$

Where $E_b = \frac{e^2}{\epsilon a_B}$ is the exciton binding energy which in monolayer TMDCs is $\sim 0.5 - 1$ eV from *ab initio* calculations [S4, S5, S12]. *Ab initio* calculations also give a Bohr radius $a_B = 1$ nm [S12]. Taking $E_b = 0.5$ eV, we find $J \sim 1$ eV.

## S2. Strain induced in-plane Zeeman field

When an in-plane uniaxial tensile strain breaks the three fold rotational symmetry, $J^{SR}_{k=0}$ is allowed to have finite value. This will generate a **k**-independent term in the valley coupling, i.e. an effective in-plane Zeeman field $J_0$ on the valley pseudospin,

$$J_0 \equiv J^{SR}_{k=0} = \sum_{G \neq 0} \frac{V(G)}{A} \left( \sum_q \psi(-q) \langle u_{-K+q,v,\downarrow} | e^{-iG \cdot r} | u_{-K+q,c,\downarrow} \rangle \right)$$

$$\times \left( \sum_{q'} \psi(q') \langle u_{K+q',c,\uparrow} | e^{iG \cdot r} | u_{K+q',v,\uparrow} \rangle \right)$$

$$\approx |\psi(r_{eh} = 0)|^2 \sum_{G \neq 0} V(G) \langle u_{-K,v,\downarrow} | e^{-iG \cdot r} | u_{-K,c,\downarrow} \rangle \langle u_{K,c,\uparrow} | e^{iG \cdot r} | u_{K,v,\uparrow} \rangle.$$

Since $G$ is large, the screening effect is negligible so $V(G) = \frac{2\pi e^2}{\epsilon G}$. Thus $J_0 \approx \frac{2\pi e^2}{\sqrt{3} \epsilon K} |\psi(r_{eh} = 0)|^2 \times f_s$, with $f_s \equiv \sum_{G \neq 0} \frac{\sqrt{3}K}{G} \langle u_{-K,v,\downarrow} | e^{-iG \cdot r} | u_{-K,c,\downarrow} \rangle \langle u_{K,c,\uparrow} | e^{iG \cdot r} | u_{K,v,\uparrow} \rangle$.

$f_s$ can be evaluated from the *ab initio* wavefunctions. We performed the *ab initio* calculations using the VASP package [S7] in conjunction with the ABINIT package [S11]. We first used VASP to relax the atomic positions in a strained cell and then used the obtained structure parameters to calculate the wavefunctions by ABINIT. In all calculations projector-augmented wave method [S8] was used in combination with the Perdew-Burke-Ernzerhof exchange-correlation functional [S9]. In the VASP calculations, the cutoff energy of the plane-wave basis is set to 370 eV with a convergence criterion of $10^{-6}$ eV. The **Γ**-centered **k** mesh is $10 \times 10 \times 1$ and layer separation is larger than 15 Å. Uniaxial strain is applied in the $\hat{x}$ (armchair) or $\hat{y}$ (zigzag) directions respectively (see Fig. 2c in the main text). Taking the 1% strain in $\hat{x}$ direction for example, in the strained unit cell the distance between the nearest two W atoms in $\hat{x}$ direction $d^x_{W-W} = 1.01 \times \sqrt{3}a$, and the W-W distance in $\hat{y}$ direction was kept unchanged, i.e. $d^y_{W-W} = a$. Here $a = 3.310$ Å is the lattice constant of the unstrained WSe$_2$ monolayer [S10]. For the unit cell in each strained case, the atomic positions are relaxed until the force on each atom is less than 0.005 eV/Å. We obtain $f_s$ as below

|  | 0% | 1% | 2% | 3% |
|---|---|---|---|---|
| Strain along $\hat{x}$ | 0 | -0.0726 | -0.1366 | -0.1912 |
| Strain along $\hat{y}$ | 0 | 0.0773 | 0.1567 | 0.2385 |

With a moderate strain of 1%, we find $J_0 = 6$ meV,

The strain induced valley coupling of excitons can be written as $\mathbf{J}_0 \cdot \hat{\boldsymbol{\sigma}}$, where $\mathbf{J}_0 \equiv (\text{Re}(J_0), -\text{Im}(J_0))$, and $\hat{\boldsymbol{\sigma}}$ is the valley pseudospin. The exciton dispersion is then $\frac{\hbar^2 k^2}{2M_0} \pm$ $|J_0 - \frac{k}{K}Je^{-2i\theta}|$. In the light cone, the dispersion can be well approximated by $E(\mathbf{k}) = \pm(J_0 - \frac{k}{K}J\cos 2\theta)$. Both the upper and lower branches correspond to saddle points with linear dispersion around $\mathbf{k} = 0$.

## S3. Exchange interaction in the negatively charged trion X-

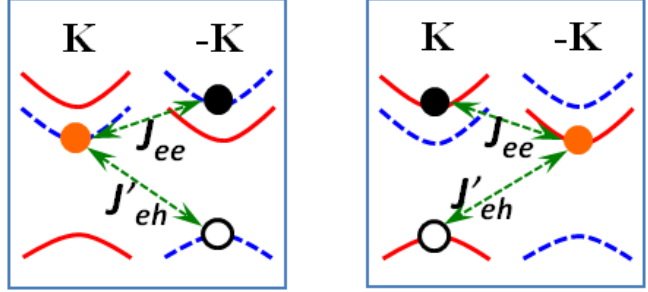

Fig. S2 The exchange energy between the excess electron (orange dot) and the recombining e-h pair (black and white circles), in the two configurations of X-.

When the exciton binds an excess electron to form a trion, we also need to take into account the exchange energy between the excess electron and the recombining e-h pair (see Fig. S2). In the four ground state configurations (c.f. Fig. 3 of main text), such an exchange energy is present only for the two shown in Fig. S2. The magnitude of the exchange energies are

$$J_{ee} \approx -f_{ee}\int d\mathbf{r}_1 d\mathbf{r}_2 \psi^*_{\mathbf{K}c\downarrow}(\mathbf{r}_1)\psi^*_{-\mathbf{K}c\downarrow}(\mathbf{r}_2)V(\mathbf{r}_1-\mathbf{r}_2)\psi_{\mathbf{K}c\downarrow}(\mathbf{r}_2)\psi_{-\mathbf{K}c\downarrow}(\mathbf{r}_1)$$
$$\approx -f_{ee}V(\mathbf{K})|\langle u_{-\mathbf{K}c\downarrow}|u_{\mathbf{K}c\downarrow}\rangle|^2,$$
$$J'_{eh} \approx f_{eh}\int d\mathbf{r}_1 d\mathbf{r}_2 \psi^*_{\mathbf{K}c\downarrow}(\mathbf{r}_1)\psi_{-\mathbf{K}v\downarrow}(\mathbf{r}_1)V(\mathbf{r}_1-\mathbf{r}_2)\psi^*_{-\mathbf{K}v\downarrow}(\mathbf{r}_2)\psi_{\mathbf{K}c\downarrow}(\mathbf{r}_2)$$
$$\approx f_{eh}V(\mathbf{K})|\langle u_{-\mathbf{K}v\downarrow}|u_{\mathbf{K}c\downarrow}\rangle|^2.$$

Here $f_{ee}$ ($f_{eh}$) corresponds to the probability of finding the electron-electron (electron-hole) in the same position. Because of the Coulomb repulsion (attraction) between the two electrons (electron and hole), $f_{ee}$ is much smaller than $f_{eh}$, and therefore we expect $J'_{eh}$ dominates over $J_{ee}$. $f_{eh} \sim a_{eh}^{-2}$, where $a_{eh}$ is the distance of the excess electron from the hole. From the *ab initio* Bloch wave function, we find $|\langle u_{-\mathbf{K}v\downarrow}|u_{\mathbf{K}c\downarrow}\rangle|^2 \approx 0.1$. Thus, $J'_{eh} \approx 0.1V(\mathbf{K})/a_{eh}^2 \sim \frac{0.2\pi}{Ka_{eh}}\frac{e^2}{\varepsilon a_{eh}}$. A numerical calculation for X- in monolayer TMDCs finds that the distances of the hole from the two electrons are respectively $\sim a_B$ and $\sim 2.5a_B$, where $a_B = 1$ nm is the Bohr radius of neutral exciton $X_0$ [S3]. Taking $a_{eh} = 2$ nm, we estimate $J'_{eh} \sim 6$ meV. This exchange energy corresponds to the trion splitting $\delta$ at the Dirac point (c.f. main text).

The gap opening of the Dirac cone gives rise to large Berry curvature for X- with center-of-mass wave vector $\pm\mathbf{K} + \mathbf{k}$ which is calculated to be

$$\Omega(\mathbf{k}) = -\hat{\sigma}_z \frac{k^2(k+2k_{\text{TF}})}{(k+k_{\text{TF}})^3}\frac{2J^2}{K^2\delta^2}\left(1 + \frac{4k^4J^2}{(k+k_{\text{TF}})^2K^2\delta^2}\right)^{-\frac{3}{2}}.$$

Here $k_{\text{TF}}$ is the 2D Thomas-Fermi screening wave vector. This Berry curvature is independent on the excess electron spin $\hat{s}_z$, but depends on the valley pseudospin $\hat{\sigma}_z$ of the recombining e-h pair, as shown in Fig. 3 of the maintext. A finite value of $k_{\text{TF}}$ leads to $\Omega(\mathbf{k}=0)=0$, as illustrated in Fig. 4 (b) of the maintext.